\documentclass[11pt]{article}
\usepackage{hyperref}

\usepackage{amsmath,amsfonts,amssymb}
\usepackage{graphicx}
\usepackage{epsfig,epsf}

\def\bar {\overline}

\def\be {\begin{equation}}
\def\ee {\end{equation}}
\def\beq {\begin{equation}}
\def\eeq {\end{equation}}
\def\bea {\begin{eqnarray}}
\def\eea {\end{eqnarray}}

\def\bra {\langle}
\def\ket {\rangle}

\def\beq{\begin{equation}}
\def\eeq{\end{equation}}
\def\barr{\begin{array}}
\def\earr{\end{array}}

\def\opcit(#1){ {\em op. cit.}, #1}

\def\issue(#1,#2,#3){#1, #2 (#3)} 

\def\APP(#1,#2,#3){Acta Phys.\ Polon.\ \issue(#1,#2,#3)}
\def\ARNPS(#1,#2,#3){Ann.\ Rev.\ Nucl.\ Part.\ Sci.\ \issue(#1,#2,#3)}
\def\CPC(#1,#2,#3){Comp.\ Phys.\ Comm.\ \issue(#1,#2,#3)}
\def\CIP(#1,#2,#3){Comput.\ Phys.\ \issue(#1,#2,#3)}
\def\EPJC(#1,#2,#3){Eur.\ Phys.\ J.\ C\ \issue(#1,#2,#3)}
\def\EPJD(#1,#2,#3){Eur.\ Phys.\ J. Direct\ C\ \issue(#1,#2,#3)}
\def\IEEETNS(#1,#2,#3){IEEE Trans.\ Nucl.\ Sci.\ \issue(#1,#2,#3)}
\def\IJMP(#1,#2,#3){Int.\ J.\ Mod.\ Phys. \issue(#1,#2,#3)}
\def\JHEP(#1,#2,#3){J.\ High Energy Physics \issue(#1,#2,#3)}
\def\JPG(#1,#2,#3){J.\ Phys.\ G \issue(#1,#2,#3)}
\def\MPL(#1,#2,#3){Mod.\ Phys.\ Lett.\ \issue(#1,#2,#3)}
\def\NP(#1,#2,#3){Nucl.\ Phys.\ \issue(#1,#2,#3)}
\def\NIM(#1,#2,#3){Nucl.\ Instrum.\ Meth.\ \issue(#1,#2,#3)}
\def\PL(#1,#2,#3){Phys.\ Lett.\ \issue(#1,#2,#3)}
\def\PRD(#1,#2,#3){Phys.\ Rev.\ D \issue(#1,#2,#3)}
\def\PRL(#1,#2,#3){Phys.\ Rev.\ Lett.\ \issue(#1,#2,#3)}
\def\SJNP(#1,#2,#3){Sov.\ J. Nucl.\ Phys.\ \issue(#1,#2,#3)}
\def\ZPC(#1,#2,#3){Zeit.\ Phys.\ C \issue(#1,#2,#3)}

\begin{document}

\begin{center}
 {\Large\bf{
Two-Higgs doublet models confront the naturalness problem}}

\vspace{5mm}

Indrani Chakraborty \footnote{indrani300888@gmail.com} 
and
Anirban Kundu \footnote{anirban.kundu.cu@gmail.com}

\vspace{3mm}
{\em{Department of Physics, University of Calcutta, \\
92 Acharya Prafulla Chandra Road, Kolkata 700009
}}
\end{center}
\begin{abstract}
The conjecture that some unknown symmetry is responsible for keeping the Higgs boson 
light at 125 GeV does not hold for the Standard Model, where the coefficient of the 
quadratic divergence of Higgs boson self-energy is far from zero. We show that such 
a cancellation can be achieved in two-Higgs doublet models, by virtue of which all the 
scalars remain at the electroweak scale and the naturalness problem is avoided. 
We explore the consequences of such cancellations in different two-Higgs doublet 
models with no flavour-changing neutral current, and show that the parameter space 
becomes tightly constrained; in particular, the ratio of two vacuum expectation values, $\tan\beta$,
no longer remains a free parameter but turns out to be a function of the quartic couplings.

\end{abstract}

\date{\today}





\section{Introduction}

The two-Higgs doublet models (2HDM) \cite{Branco:PhysRept} are one of the most widely investigated 
scenarios that go beyond the Standard Model (SM). Any 2HDM consists of five physical scalars: two 
CP-even neutral $h$ and $H$, one CP-odd neutral $A$, and two charged bosons $H^\pm$. The CP
quantum numbers are, of course, assigned with the assumption that the scalar potential is CP conserving 
and hence the mass eigenstates are also CP eigenstates. However, a generic 2HDM suffers from 
large flavour-changing neutral currents (FCNC); to prevent this, one invokes the Glashow-Weinberg-Paschos 
theorem \cite{Glashow:1976nt,Paschos:1976ay}. The theorem states that there will be no tree-level FCNC 
if all right-handed fermions of a given electric charge couple to only one of the doublets. This can be 
achieved in 2HDMs by introducing discrete symmetries for fermions or scalars.   

Let us denote the two doublets by $\Phi_1$ and $\Phi_2$, and invoke a $Z_2$ symmetry $\Phi_1\to
-\Phi_1$, $\Phi_2\to \Phi_2$. There are four types of 2HDM, depending on the 
transformation of the fermions under this $Z_2$, for which there will be no tree-level FCNC. They are:
(i) Type I, for which all fermions couple with $\Phi_2$ and none with $\Phi_1$; (ii) Type II, for which 
up-type quarks couple to $\Phi_2$ and down-type quarks and charged leptons couple to $\Phi_1$ 
(this is the type that is embedded in the minimal supersymmetric SM (MSSM) and hence has received 
the most attention); (iii) Type Y (sometimes called Type III or Flipped), for which up-type quarks and
charged leptons couple to $\Phi_2$ and down-type quarks couple to $\Phi_1$, and (iv) Type X 
(sometimes called Type IV or Lepton-specific), for which all charged leptons couple to $\Phi_1$ and 
all quarks couple to $\Phi_2$. Apart from these four, there are other 2HDMS with tree-level FCNC 
which can be kept under control \cite{bgl} and which lead to interesting phenomenology \cite{bgl2,
dipankar} which we will not pursue in this paper. 

The facts that the SM can be an effective theory valid all the way up to the Planck scale 
($M_{Pl}= 1/\sqrt{G_N} \sim 10^{19}$ GeV) and there 
is no symmetry protecting the scalar masses lead to the naturalness problem: why the Higgs boson mass 
is of the order of the electroweak scale and not driven by the radiative corrections to the Planck scale.
As we do not yet know, from experiments, of any symmetry that can protect the scalar mass, we will 
take a bottom-up approach, first suggested by Veltman \cite{veltman}, that due to some yet-to-be-discovered
symmetry, the radiative corrections to the scalar mass either vanish or are kept at a manageable 
level. This is popularly known as the Veltman condition (VC). 

The Veltman condition is not an anthropic principle. Rather, it gives a concrete guideline of how the 
SM extensions should look like if we are to address the naturalness problem. (Exact supersymmetry is 
one such guideline: a superparticle loop with an opposite sign to the corresponding particle loop, 
hence cancelling the divergence.) In the SM with a scalar 
potential 
\be
V(\Phi) = -\mu^2\Phi^\dag\Phi + \lambda (\Phi^\dag\Phi)^2 \,,
\label{sm-pot}
\ee
and the vacuum expectation value (VEV) given by $\bra \Phi \ket = v/\sqrt{2}$, the one-loop quadratically
divergent correction to the Higgs self-energy is \cite{veltman}
\be
\delta m_h^2 = \frac{\Lambda^2}{16\pi^2} \left(6\lambda + \frac34 g_1^2 + \frac94 g_2^2
- 6 g_t^2\right)\,,
\label{smvc}
\ee
where $g_1$ and $g_2$ are the $U(1)_Y$ and $SU(2)_L$ gauge couplings, 
and $g_t = \sqrt{2}m_t/v$ is the top quark Yukawa coupling. Contributions from other fermions 
are negligible. We use 
the cut-off regularization, $\Lambda$ being the cutoff scale. This is not a Lorentz invariant
regularization but has the nice feature of separating the 
quadratic and the logarithmic divergences. Dimensional 
regularization does not discriminate between these two, and one gets a
slightly different correction \cite{einhorn}, which includes all the divergences lumped into 
the ubiquitous $1/\epsilon$:
 \be
\delta m_h^2 \propto \frac{1}{\epsilon} \left(6\lambda + \frac14 g_1^2 + \frac34 g_2^2
- 6 g_t^2\right)\,.
\ee
As our goal is to cancel the strongest divergence, we will use the cut-off regularization.
There are further quadratic divergences coming from two-loop 
diagrams, but they are suppressed from one-loop contributions by a factor of 
$\ln(\Lambda/\mu)/16\pi^2$, where $\mu$ is the regularization scale, and is in general small 
and under control. This correction shifts the unphysical bare mass parameter to the physical 
pole mass of 125 GeV:
\be
m_h^2 = {m_h^{\rm bare}}^2 + \delta m_h^2\,.
\ee
Thus, $\delta m_h^2/m_h^2$ is a measure of fine-tuning of the theory. 
The strict VC implies $\delta m_h^2 = 0$. We may say that the naturalness problem has been 
avoided if $|\delta m_h^2| \leq m_h^2$ \footnote{This means no fine-tuning. If one is ready to accept 
a certain amount of fine-tuning, say 1\%, the condition would have been $|\delta m_h^2| 
\leq 100 m_h^2$.}, 
which translates into
\be
\left\vert m_h^2 + 2m_W^2 + m_Z^2 - 4m_t^2\right\vert \leq \frac{16\pi^2}{3} \frac{v^2}{\Lambda^2} m_h^2\,.
\label{smvc2}
\ee
This inequality is clearly not satisfied in the SM for $v^2/\Lambda^2 \leq 0.1$, or $\Lambda \geq 760$ GeV,
and onset of NP at such a low scale is almost ruled out by the LHC. 

It is clear from Eq.\ (\ref{smvc}) that some positive contribution to $\delta m_h^2$ is needed to
offset the large negative contribution of $-6g_t^2$. Thus, the minimal extension of the SM that 
can satisfy the VC must be bosonic in nature. The easiest option is to introduce new scalars, 
with the rider that it is imperative to keep these new scalars light too; so simultaneous solution to 
a set of VCs is required. One might also introduce 
new gauge bosons that couple to the SM Higgs, but more scalars are anyway needed to give mass
to these gauge bosons in a gauge-invariant way. 

With new scalars, there are two ways to solve the VCs. Some of the scalar couplings may be negative.
This gives a negative contribution to the VC, akin to the fermion loops. However, one must take 
into account the stability of the potential at all energy scales. The second one is to allow explicit 
couplings to the fermions for all the scalars. Apart from the 2HDM Type-I, all other 2HDMs have 
fermionic couplings for both $\Phi_1$ and $\Phi_2$. 

The role of the VC in exploring possible directions for new physics has been well explored in the 
literature. However, the minimal extension, with one or more singlet scalars, received the most attention
\cite{aksrc,drozd,indrani,bazzocchi}. Such singlets are interesting as possible cold dark matter candidates
if they do not mix with the doublet Higgs. In that case, one is forced to introduce some new fermion
multiplets, vectorial in nature, to address the fine-tuning of the singlets. A model with a complex scalar 
triplet has also been investigated \cite{indrani-2}, which is relevant in the context of type-2 see-saw 
mechanism and leptogenesis (even though the minimal version of the triplet model does not work properly 
for the latter). The attractive point about the triplet is that no extra fermions are
needed; the neutrinos couple with the triplet scalars and can potentially solve the fine-tuning 
problem, at the same time ensuring a small neutrino mass. 

The 2HDMs share the same property of requiring no 
extra fermions, and are therefore interesting as a minimal extension to the SM to successfully address 
the fine-tuning problem. Consequences of applying the VC to 2HDMs were discussed in Ref.\
\cite{ma:0101355} and then later discussed in more detail in Ref.\ \cite{bohdan:0910.4068}. 
Of course, the Higgs boson was not discovered then and the parameter space was much larger, 
but we would still like to refer the reader to \cite{bohdan:0910.4068} for some points not 
covered in detail here, like the possibility of CP violation in the scalar sector, or 
enhancing the 2HDM with one more singlet to generate a cold dark matter candidate. 

Except for Type I 2HDM, fermionic couplings exist for all the scalars. 
Even without fermions,
the VC could have been satisfied with some negative scalar quartic couplings, but the stability condition 
of the scalar potential rules that out, as we will see later. It will also be shown that the 
imposition of the VC puts a significant constraint on the 
parameter space. In particular, $\tan\beta$, the ratio of the VEVs of two 
scalars, is no longer a free parameter. Later we will try to quantify this statement. 
We will also show that only Type-II and Flipped 2HDMs can remain valid up to the Planck scale 
if the couplings are to satisfy the VCs and also respect the stability criteria of the potential. 

The paper is arranged as follows. In Section II, we display the scalar potential and Yukawa interactions of 
different 2HDMs and the stability conditions; we also formulate the Veltman conditions. 
The analysis is in Section III, and Section IV summarizes and concludes the paper. 
The one-loop renormalization group equations (RGE) for all the couplings are listed in Appendix A.

\section{2HDM in brief}

We will follow the notations and conventions of Ref.\ \cite{Branco:PhysRept} and confine ourselves only
to the four types of 2HDM with no tree-level FCNC. This can be achieved with the following discrete 
symmetries (we show only those fields that flip sign under $Z_2$):
\begin{itemize}
 \item Type I: $\Phi_1 \to -\Phi_1$;
 \item Type II: $\Phi_1 \to -\Phi_1$, $d^i_R\to -d^i_R$, $e^i_R\to -e^i_R$;
 \item Lepton specific: $\Phi_1 \to -\Phi_1$, $e^i_R\to -e^i_R$;
 \item Flipped: $\Phi_1 \to -\Phi_1$, $d^i_R\to -d^i_R$\,,
\end{itemize}
where $i$ is the generation index. Both the scalar doublets $\Phi_1$ and $\Phi_2$ 
have hypercharge $+1$, and the lower components, which are electrically neutral, get 
nonzero VEV:
\be
\bra\Phi_1\ket = \frac{1}{\sqrt{2}} \begin{pmatrix} 0 \cr v_1 \end{pmatrix}\,, 
\bra\Phi_2\ket = \frac{1}{\sqrt{2}} \begin{pmatrix} 0 \cr v_2 \end{pmatrix}\,, 
\ee
with $\tan\beta = v_2/v_1$ (without loss of generality, $\beta$ can be taken to lie 
in the first quadrant, so that both $v_1,v_2 > 0$) 
and $m_W = \frac12 g_2 \sqrt{v_1^2+v_2^2}$.
The CP-conserving scalar potential can be written as
\bea
 V &=&  m_{11}^2 \Phi_1^\dag\Phi_1 + m_{22}^2 \Phi_2^\dag\Phi_2 - m_{12}^2 \left(\Phi_1^\dag\Phi_2 
  + \Phi_2^\dag\Phi_1\right)\nonumber\\
  && + \frac12\lambda_1\left(\Phi_1^\dag\Phi_1\right)^2+\frac12\lambda_2\left(\Phi_2^\dag\Phi_2\right)^2
     + \lambda_3 \left(\Phi_1^\dag\Phi_1\right)\left(\Phi_2^\dag\Phi_2\right) \nonumber\\
&&   + \lambda_4 \left(\Phi_1^\dag\Phi_2\right)\left(\Phi_2^\dag\Phi_1\right)+
\frac12\lambda_5\left[\left(\Phi_1^\dag\Phi_2\right)^2 + \left(\Phi_2^\dag\Phi_1\right)^2\right]\,,
\eea
where $m_{12}^2$ softly breaks the $Z_2$ symmetry \footnote{One can also have quartic terms in the 
scalar potential that break $Z_2$. These terms have odd numbers of $\Phi_1$ and $\Phi_2$ and 
hence do not conribute to the quadratic divergences at the one-loop level.}. 
The condition for extremum of the potential is
\bea
m_{11}^2 - m_{12}^2 \, \tan\beta  + \frac12 \lambda_1 v_1^2 + \frac12 
\left(\lambda_3+\lambda_4+\lambda_5\right) v_2^2 &=& 0\,,\nonumber\\
m_{22}^2 - m_{12}^2 \, \cot\beta  + \frac12 \lambda_2 v_2^2 + \frac12 
\left(\lambda_3+\lambda_4+\lambda_5\right) v_1^2 &=& 0\,.
\label{minimiz}
\eea
The two CP-even neutral states $\rho_1$ and $\rho_2$, which are components of $\Phi_1$ and $\Phi_2$
respectively, are not mass eigenstates. The corresponding mass matrix can be diagonalized through 
a rotation by an angle $\alpha$, and the mass eigenstates are\footnote{One can always replace 
$\alpha$ by $\pi+\alpha$ which introduces an otherwise irrelevant overall minus sign.}
\be
h = \rho_2\cos\alpha - \rho_1\sin\alpha\,,\ \ 
H = \rho_2\sin\alpha + \rho_1\cos\alpha\,,
\ee
where $h (H)$ is the lighter (heavier) eigenstate. Note that if $\beta - \alpha = \pi/2 (0)$, $h(H)$ 
will be the SM Higgs boson, with a VEV of $v = \sqrt{v_1^2+v_2^2}$. For example, 
the $hVV^\ast$ ($HVV^\ast$) coupling is just the SM coupling times 
$\sin(\beta - \alpha)$ ($\cos(\beta - \alpha)$, and the 
$hhVV^\ast$ ($HHVV^\ast$) coupling is the SM coupling times 
$\sin^2(\beta - \alpha)$ ($\cos^2(\beta - \alpha)$), where $V$ 
is any weak gauge boson:
\bea
hW^+W^-\,\, &:& \frac{i}{2}g_2^2 v \eta_{\mu\nu} \sin(\beta - \alpha)\,,\nonumber\\
HW^+W^-\,\, &:& \frac{i}{2}g_2^2 v \eta_{\mu\nu} \cos(\beta - \alpha)\,,\nonumber\\
hhW^+W^-\,\, &:& \frac{i}{2}g_2^2 \eta_{\mu\nu} \sin^2(\beta - \alpha)\,,\nonumber\\
HHW^+W^-\,\, &:& \frac{i}{2}g_2^2 \eta_{\mu\nu} \cos^2(\beta - \alpha)\,.
\eea
The CP-odd
scalar $A$ does not couple to gauge bosons. The oblique corrections from such couplings have been 
computed in Ref.\ \cite{2hdmoblique}. 

Before we proceed any further, let us note that one should formulate the VCs for $h$ and $H$.
However, if we demand the quadratic divergences for both $h$ and $H$ to vanish, we might as well 
formulate them for $\rho_1$ and $\rho_2$. This is what we will do in our subsequent discussion,
and perform the entire analysis in terms of the couplings and not the masses. While the propagators 
are not uniquely defined in the $\Phi_1$--$\Phi_2$ basis, this does not affect our analysis as long as we
focus on purely the divergent terms.

The most generic Yukawa interactions for these four models can be written as 
\cite{Branco:PhysRept}
\be
\mathcal{L}_Y = -\sum_{j=1}^2 \left[ Y_j^d \bar{Q}_L d_R \Phi_j + Y_j^u \bar{Q}_L u_R \tilde{\Phi}_j
+ Y_j^e \bar{L}_L l_R \Phi_j + {\rm h.c.}\right] \,,
\ee
where $\tilde{\Phi}_j = i{\tau_2}\Phi_j^*$, $Q_L$, $L_L$, $d_R$, $u_R$ and $l_R$ are 
generic doublet quarks, doublet leptons, singlet down-type and singlet up-type quarks, and singlet 
charged leptons respectively. $Y_j^d$, $Y_j^u$, $Y_j^e$ are $3\times 3$ complex matrices, 
containing Yukawa couplings for the down, up, and leptonic sectors respectively. In our analysis 
we will consider only top, bottom, and $\tau$ Yukawa couplings to be nonzero. 

\subsection{Stability conditions}

The requirement that the scalar potential always remains bounded from below leads to the 
following stability conditions \cite{Branco:PhysRept}:
\be
\lambda_1\,,\lambda_2 \ge 0\,,\ \ 
\lambda_3\geq -{\sqrt{\lambda_1\lambda_2}}\,,\ \ 
\lambda_3 + \lambda_4 - \vert\lambda_5\vert \geq  -{\sqrt{\lambda_1\lambda_2}}\,.
\label{stability}
\ee
Thus, $\lambda_3$, $\lambda_4$, and $\lambda_5$ can potentially be negative. There can 
be charge-breaking or CP-breaking stable points of the potential; however, if the normal 
minimum is deeper, such stable points can at best be saddle points \cite{0406231,ivanov}. 
There can, of course, be more than one normal stable points of 2HDM \cite{ferreira}. 
The last condition shows that $\lambda_5=0$ leads to the most stable configuration for a given
set of  the other quartic couplings. We have checked, explicitly, that for all the 2HDMs 
discussed here, there is at least one CP- and charge-conserving minimum for the parameter space points
under consideration, in particular for the benchmark points mentioned later.

\subsection{Veltman conditions}

If the Yukawa couplings are neglected, the VCs for $\rho_1$ and $\rho_2$ are the same for 
all 2HDMs. The self-energy corrections are \footnote{We are not in the mass basis, so 
these are, strictly speaking, the corrections to the 2-point Green's functions.}
\be
\delta' m_{\rho_1}^2 = \frac{\Lambda^2}{16\pi^2}\left[\left(\frac94 g_2^2 + \frac34 g_1^2 
\right)+ 2 \lambda_3 + 3 \lambda_1 + \lambda_4\right] 
\equiv \frac{\Lambda^2}{16\pi^2} f'_{\rho_1}\,,
\label{Phi_1}
\ee
\be
\delta' m_{\rho_2}^2 = \frac{\Lambda^2}{16\pi^2}\left[\left(\frac94 g_2^2 + \frac34 g_1^2 
\right)+ 2 \lambda_3 + 3 \lambda_2 + \lambda_4\right]
\equiv \frac{\Lambda^2}{16\pi^2} f'_{\rho_2}\,.
\label{Phi_2}
\ee
Even if we neglect the gauge couplings, there are no solutions consistent with Eq.\
(\ref{stability}), except the trivial solution $\lambda_i = 0$. Note that there is no term 
proportional to  $\lambda_5$; the quadratically divergent contributions cancel out. 

With the introduction of the Yukawa couplings (only for $t$, $b$, and $\tau$), the corrections 
turn out to be as follows.

\begin{itemize}
 \item Type I:
  \be
 f_{\rho_1} = f'_{\rho_1}\,,\  \ \ \ 
 f_{\rho_2} = f'_{\rho_2} - 3\left({Y_2^b}\right)^2 - 3\left({Y_2^t}\right)^2 - \left({Y_2^\tau}\right)^2\,.
 \label{type_1}
 \ee
 
 \item Type II:
  \be
 f_{\rho_1} = f'_{\rho_1} - 3\left({Y_1^b}\right)^2 - \left({Y_1^\tau}\right)^2 \,,\ \ \ \ 
 f_{\rho_2}  = f'_{\rho_2} - 3\left({Y_2^t}\right)^2 \,.
 \label{type_2}
 \ee
 
 \item Lepton-specific: 
  \be
 f_{\rho_1}= f'_{\rho_1} - \left({Y_1^\tau}\right)^2\,, \ \ \ \ 
 f_{\rho_2}  = f'_{\rho_2}  - 3\left({Y_2^b}\right)^2 - 3\left({Y_2^t}\right)^2 \,.
 \label{type_ls}
 \ee
 
 \item Flipped:
\be
 f_{\rho_1} = f'_{\rho_1} - 3\left({Y_1^b}\right)^2\,,\ \ \ \ 
 f_{\rho_2} = f'_{\rho_2} - 3\left({Y_2^t}\right)^2 - \left({Y_2^\tau}\right)^2 \,.
 \label{type_f}
 \ee
 
 \end{itemize}
 
Thus, the complete one-loop quadratically divergent corrections are
\be
\delta m_{\rho_{1(2)}}^2 = \frac{\Lambda^2}{16\pi^2} f_{\rho_{1(2)}}\,,
\ee
and the strict enforcement of the VCs require $f_{\rho_1} = 0$, $f_{\rho_2} = 0$. In addition, 
if they are to hold at all energy scales, we also need $df_{\rho_1}/d(\ln q^2) = 0$, 
 $df_{\rho_2}/d(\ln q^2) = 0$.

 \subsection{RG Equations}

We would also like to see how stable the VCs are, {\em i.e.} whether they remain more or less close 
to zero or vary wildly as go up the energy scale. While stability ensures a proper solution to 
the fine-tuning problem at all energy scales, one must take into account that (i) this is only a one-loop 
analysis and therefore strict enforcement of the VC may not be possible, and (ii) we work in the weak 
basis $(\rho_1,\rho_2)$ and not the mass basis $(h,H)$, so the corrections can be linked with the 
mass eigenstates only after a proper basis rotation. As the corrections for $\rho_1$ and $\rho_2$ are 
different, the mixing angle $\alpha$ will also change with the energy scale.

Without the Yukawa couplings, the RGEs for all 2HDM would have been the same \cite{2hdmrg}. Simplified 
expressions for the one-loop RGEs, keeping only the top, bottom, and $\tau$ Yukawa couplings, are given 
in Appendix A. Detailed expressions can be found in Ref.\ \cite{Branco:PhysRept}. Note that the coupled nature 
of the RGEs for $\lambda_1$--$\lambda_4$ ensures that if one of them hits the Landau pole, all the rest
will do the same almost at the same point. The only exception is $\lambda_5$; if it is zero to start 
with, it will always remain zero. The values of all the couplings are taken to be at the weak scale 
$q^2=m_Z^2$ initially and run upwards.

\section{Analysis}

Our strategy of analysis is going to be as follows. For every 2HDM (except Type-I), we first find out the 
parameter space for the five $\lambda_i$ couplings as well as $\tan\beta$ 
for which both the VCs are satisfied at the electroweak
scale. Because of the higher-order effects and the uncertainties in the gauge and top Yukawa couplings, we 
do not expect an exact cancellation, and will settle for 
$\vert f_{\rho_1}\vert, \vert f_{\rho_2}\vert < 0.01$. Considering the magnitude of possible 
higher-order effects \footnote{The two-loop divergences are suppressed compared to the leading ones by 
a factor of $\ln(\Lambda^2/m^2)/16\pi^2$.}, we consider this limit to be a pretty conservative estimate. 

The mass term $m_{12}^2$ has nothing to do with the quadratic divergences, but it controls the physical 
scalar masses. In the $(\rho_1,\rho_2)$ basis, the CP-even mass matrix is
\be
M_{\rm{CP-even}} = \begin{pmatrix}
                    m_{12}^2\tan\beta + \lambda_1 v^2\cos^2\beta & -m_{12}^2+ \frac12(\lambda_3+\lambda_4+\lambda_5)
                    v^2\sin 2\beta \cr
                    -m_{12}^2+ \frac12(\lambda_3+\lambda_4+\lambda_5) v^2\sin 2\beta & 
                    m_{12}^2\cot\beta + \lambda_2 v^2\sin^2\beta
                   \end{pmatrix}
\ee
which follows from the minimization conditions. We choose $m_{12}$ in such a way that the lightest 
CP-even state has a mass between 123 and 127 GeV, keeping the window for possible higher-order 
corrections. We also ensure that the couplings of $h$ to fermions and gauge bosons are SM-like, 
i.e.\ $\cos^2(\alpha-\beta) \sim 0$. The charged Higgs mass is given by
\be
m^2_{H^+} = 2m_{12}^2/\sin 2\beta -\frac12\, v^2(\lambda_4 + \lambda_5)
\ee
and the CP-odd state has a mass 
\be
m^2_A = 2\, m_{12}^2/\sin 2\beta -v^2 \lambda_5\,.
\ee
This shows that in the large $\tan\beta$ limit ($\sin 2\beta \sim 0$), in which we will be interested,
$m_{12}^2$ should be positive unless $\lambda_5$ (and maybe $\lambda_4$) is large and negative. 
It can be easily checked that such large negative values of $\lambda_5$ at the electroweak scale 
lead to a further downward running of $\lambda_5$ at higher energy scales and hence make the 
potential unstable very quickly. Thus, we will always take $m_{12}^2$ to be positive in our analysis. 

That $m_{12}^2$ cannot be large and negative can also be checked from the minimization conditions 
$\partial^2 V(v_1,v_2)/\partial v_1^2 > 0$, $\partial^2 V(v_1,v_2)/\partial v_2^2 > 0$. With Eq.\
(\ref{minimiz}), this leads to a lower bound
\be
m_{12}^2\tan\beta + \lambda_1 v_1^2 > 0\,,\ \ 
m_{12}^2\cot\beta + \lambda_2 v_2^2 > 0\,.
\ee

We have also imposed the charged Higgs mass limit of $m_{H^+} > 300$ GeV for Type-II and flipped models, 
mostly coming from $b\to s\gamma$ \cite{bsgamma}. There are no such limits on the lepton-specific model 
(and also on Type-I), but the non-observation at the LHC has been taken into account. 

For all the models, we have scanned over the entire parameter space specified by the quartic couplings
$\lambda_1$--$\lambda_5$, $m_{12}$ and $\tan\beta$. We consider only those values of the aforesaid 
parameters that pass the stability criteria of the potential and also give at least one charge- and CP-conserving 
minimum. After this, we impose both the VCs, but 
not in the strictest sense; we keep a narrow window for possible two-loop and higher-order corrections.

\subsection{Type I}
 
As $\Phi_1$ does not couple to fermions, the VC for $\rho_1$, Eq.\ (\ref{type_1}), has to be satisfied 
from negative quartic couplings $\lambda_3$ and $\lambda_4$. However, it is easy to check that this 
makes the vacuum unstable. This is a generic feature over the entire parameter space and therefore we 
will not further discuss the Type-I 2HDM. Suffice it to say that the fine-tuning problem remains in Type-I 
2HDM, and probably in all 2HDMs where at least one of the doublet does not couple to the fermions. 

\subsection{Type II}

Even before we start the analysis, it is intuitively obvious that the VCs can only be satisfied in the 
large $\tan\beta$ region if the $\lambda_i$ couplings are small, as only in this region $Y^t_2$ is 
comparable in magnitude with $Y^b_1$ or $Y^\tau_1$. 
Again, in the large $\tan\beta$ region, $Y^t_2$,
which is proportional to $m_t/\sin\beta$, is more or less constant. As the other terms in $f_{\rho_2}$ 
are determined from the VC, $\vert f_{\rho_1}\vert < 0.01$, this locks the allowed values of $\lambda_2$ 
to a narrow range. 

 For our scan, we have used the following ranges, but making sure that the values are 
consistent with the stability criteria:
\be
0 \leq \lambda_1 \leq 0.4\,, \ \ 
0 \leq \lambda_2 \leq 0.4\,,\ \ 
-0.15 \leq \lambda_3 \leq 0.25\,,\ \ 
-0.15 \leq \lambda_4 \leq 0.25\,,\ \
30.0 \leq \tan\beta \leq 51.0\,.
\label{type-2-range}
\ee
We have also scanned a wide range of values for $m_{12}$, but kept $\lambda_5=0$ to 
explore the maximum region in the parameter space where the potential is stable. Note that 
in this case $m_{12}^2 > 0$. 
While our explored parameter space is not exhaustive, there are a few interesting observations.

First, $\tan\beta$ is going to be large, as expected. The lowest possible value is 31.3, for $\lambda_1
=0$, and increases almost linearly with $\lambda_1$, ending at 49.2 for $\lambda_1 = 0.35$. For any 
chosen value of $\lambda_1$, the allowed range of $\tan\beta$ is  narrow; we found it
nowhere to be more than 0.35 \footnote{This margin depends on how strictly we wish to impose the
VCs. For a strict imposition, the width goes to zero.}. This is, 
in fact, a general conclusion for type II and flipped models; if the VCs are to be satisfied, $\tan\beta$ is 
no longer an independent variable, rather it turns out to be a function of the quartic couplings. 
The Yukawa couplings are obviously fixed by the fermion masses and $\tan\beta$. 

Second, $\lambda_2$ is bound, again as expected, the weak scale value being between 0 and $0.25$. 
While this is not physically meaningful itself, let us note that this gives a large contribution 
to the (22)-element of the CP-even mass matrix, in particular for large $\tan\beta$. 

There are no such constraints on $\lambda_3$, $\lambda_4$, and $m_{12}$, except 
that the lightest CP-even scalar should be between 123 and 127 GeV, and the charged scalar should be
more than 300 GeV. Over the entire parameter space, we found $\cos^2(\alpha-\beta) < 0.006$, so that
the lightest CP-even eigenstate behaves almost entirely like the SM Higgs boson. This tells immediately 
that $\alpha$ must be very small as $\tan\beta$ is large; for Type-II 2HDM, we found $\alpha$ 
to lie between $-0.0$ and $-0.02$. 

We have also chosen a couple of benchmark points to study the model in further details 
and illustrate the salient features.

\begin{itemize}
 \item 
 For the first benchmark point (hence called BMtype2-1), we take $\lambda_1=0.0$, $\lambda_2=0.1$, 
 $\lambda_3= 0.065$, $\lambda_4=0.02$ and $\lambda_5=0$. From a very narrow allowed range, we choose 
 $\tan\beta = 33.5$. These values, and from now on all the benchmark values, are at the electroweak scale 
 $q^2=m_Z^2$. Corresponding values for the Yukawa
 couplings are $Y_1^{b} = 0.61$, $Y_1^{\tau} = 0.24$, $Y_2^{t} = 0.71$. To keep $m_{H^+} > 300$ GeV, 
 we need $m_{12} > 10$ GeV. The RG evolution of the couplings show that the model remains perturbative 
 almost till the Planck scale for the first benchmark.
 
  \begin{figure}[!htbp]
  
\begin{center}
{
\rotatebox{0}{\epsfxsize=7.5cm\epsfbox{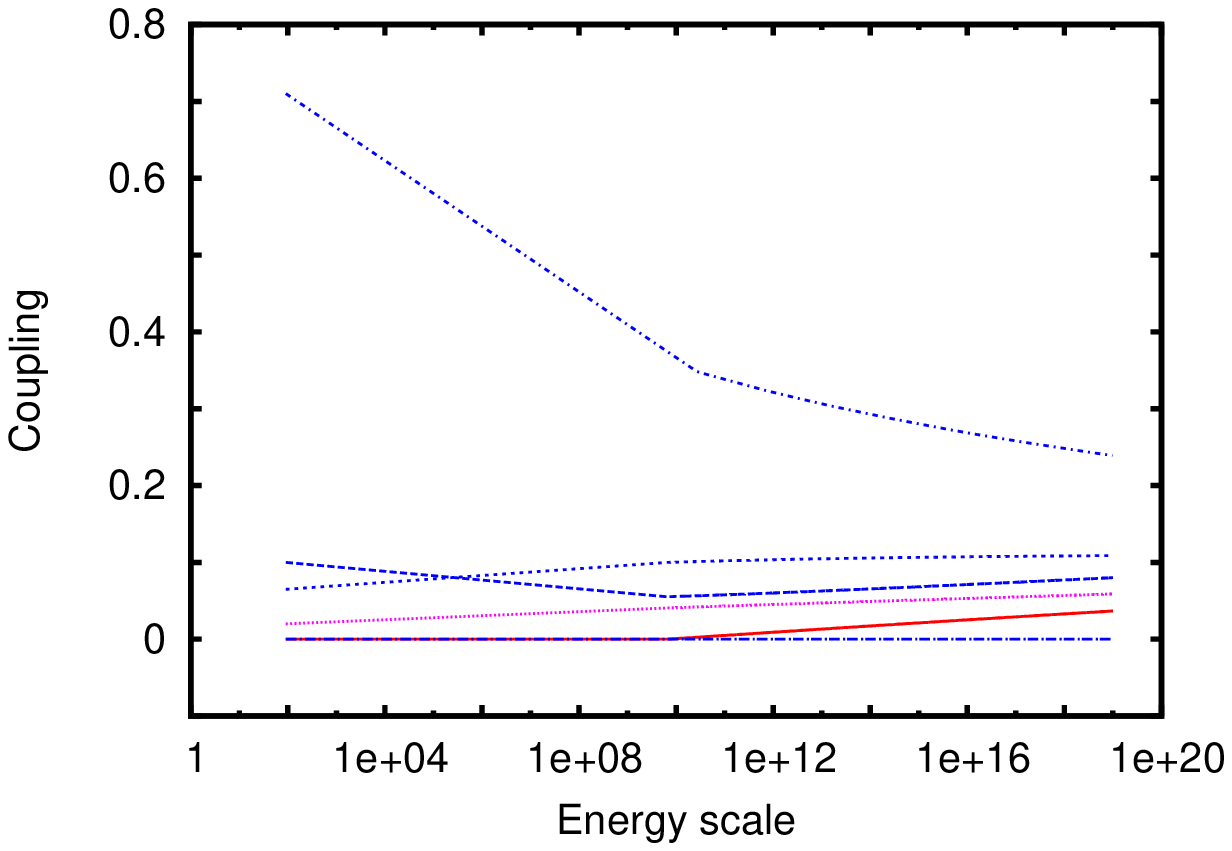}}
}
{
\rotatebox{0}{\epsfxsize=7cm\epsfbox{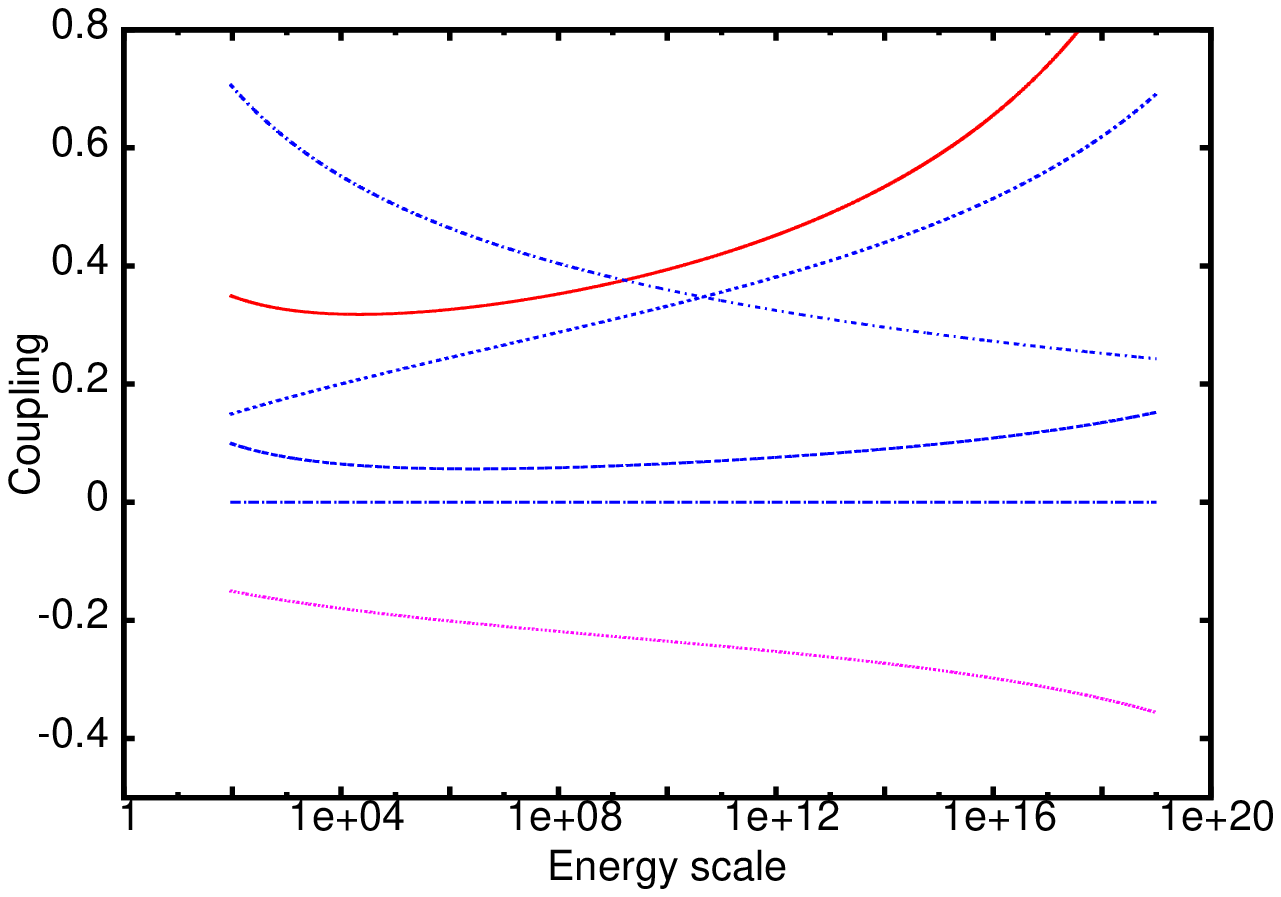}}
}
\end{center}

\caption{Running of couplings for the first and second benchmark points in Type-II 2HDM. 
The line held steady at 0 is for $\lambda_5$ (dash-dot), the slowly falling line is for $Y^t_2$ (short dash-dot). 
The other four lines are, from top to bottom at the right-edge, (i) $\lambda_3$ (short dash), $\lambda_2$ 
(long dash), $\lambda_4$ (dot) and $\lambda_1$ (solid) for the left plot (BMtype2-1), 
(ii) $\lambda_1$ (solid), $\lambda_3$ (short dash), $\lambda_2$ (long dash) and 
$\lambda_4$ (dot) for the right plot (BMtype2-2).}
\label{fig:benchmark1}
\end{figure}

 \item The second benchmark point (BMtype2-2) is specified by $\lambda_1=0.35$, $\lambda_2=0.10$, $\lambda_3=0.15$, 
 $\lambda_4= -0.15$, and $\lambda_5=0$; $\tan\beta$ is constrained at $46.0$, and $m_{12} > 10$ GeV. 
 The Yukawa couplings are $Y_1^{b} = 0.84$, $Y_1^{\tau} = 0.33$, and $Y_2^{t} = 0.71$. Note that all the couplings 
 remain perturbative till the Planck scale for this benchmark. 
 
 \end{itemize}

 \begin{figure}[!htbp]
  \begin{center}
{
\epsfxsize=7cm\epsfbox{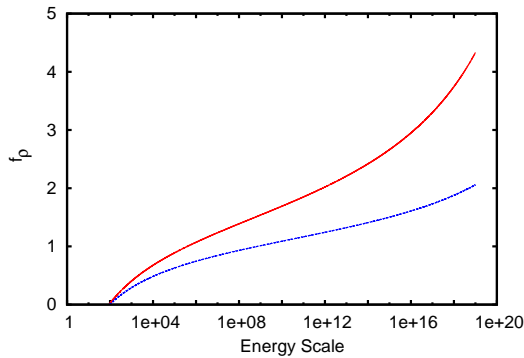}
}
\end{center}
\caption{Variation of $f_{\rho_1}$ (red, solid) and $f_{\rho_2}$ (blue, dashed) with energy for Type-II 2HDM.}
\label{benchmark2_VC}
\end{figure}
 
 The Yukawa coupling $Y^t_2$ goes down with energy (see Fig.\ \ref{fig:benchmark1} or Eq.\ (\ref{rge2})), 
while the scalar quartics either remain stable or increase. This destabilizes the VC for $\rho_2$. The destablization is 
more for higher values of the $\lambda_i$ couplings. We show, in Fig.\ 
\ref{benchmark2_VC}, the behaviour of $f_{\rho_1}$ and $f_{\rho_2}$ for BMtype2-2; 
the former is more under control than the latter,
as expected. While it may appear that the VCs are satisfied only 
at the electroweak scale but goes unmanageably out of control at higher $q^2$, we again stress that 
(i) this is only a one-loop calculation and unless the higher-order terms are computed, there is no way 
to know whether the behaviour is stable; and (ii) there may be some new physics at an intermediate scale that 
modifies the VCs.  Also note that the fine-tuning of the physical scalar masses will be different, the 
corresponding functions $f_h$ and $f_H$ being some 
combinations of $f_{\rho_1}$ and $f_{\rho_2}$.


 \subsection{Flipped}

The flipped 2HDM is very similar to Type-II, in fact, if we neglect the $\tau$ Yukawa 
coupling, they are identical. So we expect almost the same behaviour: a very tight 
constraint on $\lambda_2$, which turns out to be below $0.05$, and range for 
$\tan\beta$ is from 40.3 to 50.5. We take $m_{12} > 10$ GeV to keep $m_{H^+} > 300$ GeV.  
The Higgs mixing angle $\alpha$ lies between $-0.025$ and $-0.019$. 

Our scan range is
\be
0.1 \leq \lambda_1 \leq 0.4\,, \ \ 
0 \leq \lambda_2 \leq 0.2\,,\ \ 
-0.05 \leq \lambda_3 \leq 0.2\,,\ \ 
-0.05 \leq \lambda_4 \leq 0.25\,,
\label{type-ls-range}
\ee
and $\lambda_5 = 0$. The allowed range of $\tan\beta$ is $40.3 < \tan\beta < 50.5$.  
 
\begin{itemize}
\item For the first benchmark (BMflip-1), we take 
$\lambda_1 = 0.35$, $\lambda_2 = 0.05$, $\lambda_3=0.18$, $\lambda_4=-0.045$, 
$\lambda_5= 0$ and $\tan\beta= 49.0$. 
The Yukawa couplings are, $Y_1^{b} = 0.90$, $Y_2^{\tau} = 0.007$, and $Y_2^{t} = 0.71$. 
Note that all the 
couplings are well-behaved till the Planck scale.
 
\begin{figure}[!htbp]
 \begin{center}
{
\epsfxsize=7cm\epsfbox{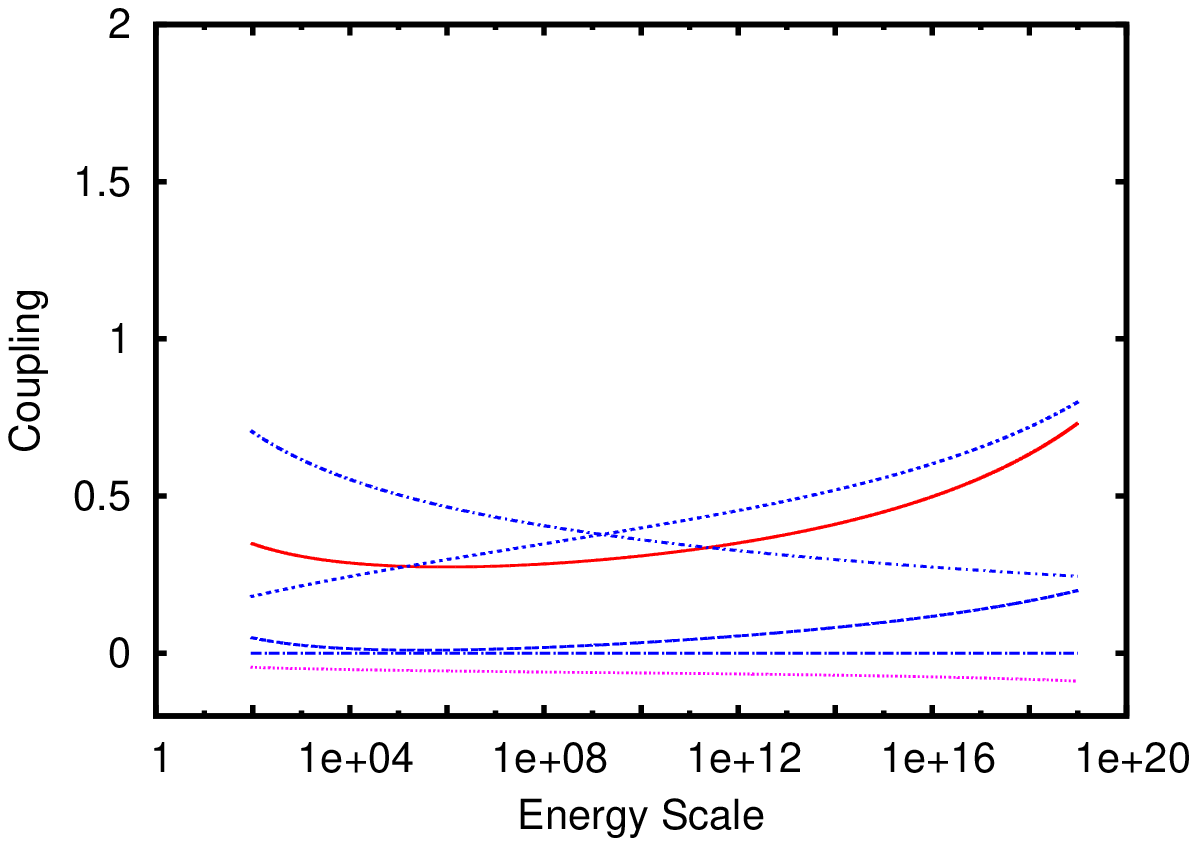}
\epsfxsize=7cm\epsfbox{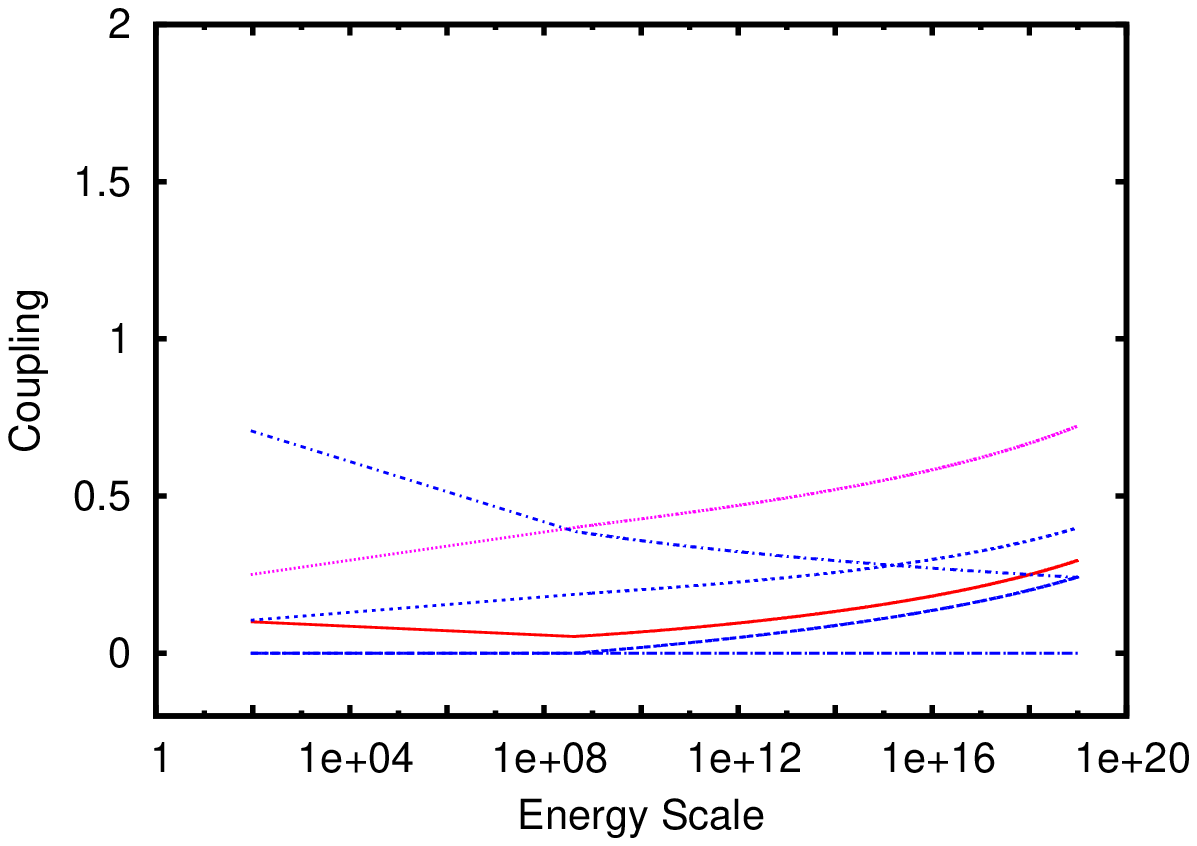}
}
\end{center}

\caption{
Running of couplings for the first and second benchmark points in the flipped 2HDM. 
The line held steady at 0 is for $\lambda_5$ (dash-dot), the slowly falling line is for $Y^t_2$ (short dash-dot). 
Among the others, for BMflip-1 (left plot), lines at the left edge are for  
$\lambda_1$ (solid), $\lambda_3$ (short dash), $\lambda_2$ (long dash), and $\lambda_4$ (dot) 
(from top to bottom) respectively. For BMflip-2 (right plot),
lines at the left edge, from top to bottom, are for $\lambda_4$ (dot), $\lambda_3$ (short dash), $\lambda_1$ 
(solid) and $\lambda_2$ (long dash) respectively. }
\label{fig:benchmark3}
\end{figure} 

\item The second benchmark (BMflip-2) is taken at
$\lambda_1=0.1$, $\lambda_2=0.0$, $\lambda_3= 0.19$, $\lambda_4=0.07$, $\lambda_5=0.0$ and 
  $\tan\beta = 42.5$. The Yukawa couplings are, $Y_1^{b} = 0.778$, $Y_2^{\tau} = 0.007$, $Y_2^{t} = 0.7075$.
Again, the model is stable till the Planck scale.

\end{itemize}

\begin{figure}[!htbp]
  \begin{center}
{
\epsfxsize=7cm\epsfbox{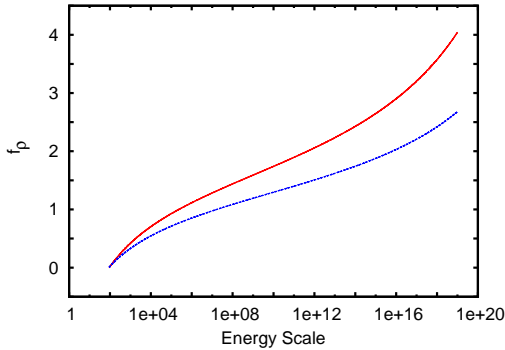}
}
\end{center}
\caption{Variation of $f_{\rho_1}$ (red, solid) and $f_{\rho_2}$ (blue, dashed) with energy for flipped 2HDM.}
\label{fig:benchmark1_VC}
\end{figure}

The stability of $f_{\rho_1}$ and $f_{\rho_2}$ is shown in Fig.\ \ref{fig:benchmark1_VC} for BMflip-1.

\subsection{Lepton specific}

Lepton-specific 2HDM also needs large values of $\tan\beta$ because 
the $\tau$-Yukawa coupling has to play the deciding role in the VC of $\rho_1$. 
In fact, it is even larger than that needed for Type-II and flipped 2HDM, as $m_{\tau}
< m_b$. 

Our scan range is
\be
0 \leq \lambda_1 \leq 0.4\,, \ \ 
0.0 \leq \lambda_2 \leq 0.4\,,\ \ 
-0.05 \leq \lambda_3 \leq 0.4\,,\ \ 
-0.05 \leq \lambda_4 \leq 0.4\,.
\label{lepsp-range}
\ee
We find $141 < \tan\beta < 221$ with $\lambda_5 = 0$, and $-0.008 <\alpha < -0.004$. 
Such large values of $\tan\beta$ 
is known to be problematic for possible non-perturbative behavious of the Yukawa couplings. 
As the Yukawa couplings have a negative pull on the quartics, large values for them 
tend to make the quartic couplings negative very quickly, making the potential unstable. 
In fact, for the entire allowed parameter space, such instability occurs below 1 TeV. 
Thus, one is forced to predict some cutoff at 1 TeV or below, above which a new dynamics 
takes over.

 \section{Summary}
 
We have discussed the fine-tuning problem in the context of 2HDMs. In particular, we try to find the 
parameter space in several 2HDMs where the quadratic divergences of the scalar masses are under control. 

We have kept ourselves confined only among those models that do not have any tree-level FCNC, and explore 
the possible parameter space where the Veltman conditions are satisfied for all the scalars. This is helped 
by the fermionic couplings of the scalars; Type-I 2HDM does not possess this feature, $\Phi_1$ does not have 
any fermionic coupling, so a successful cancellation of the $\Lambda^2$ terms is never possible, even 
if we take some quartic scalar couplings to be negative. We emphasize that our results are valid only for 
standalone 2HDMs, not when they are embedded in a larger theory.

In contrast with the SM (or the Type-I 2HDM), it is possible to find a parameter space for all the other
three 2HDMs where the VCs are satisfied. This may not seem very surprising, as we have more parameters 
to play with, but note that we also have a large number of constraints: the stability of the potential, 
the existence of a CP-even minimum,
the lightest CP-even neutral scalar close to 125 GeV, the constraint on the charged Higgs mass, the 
necessity of making the 125 GeV scalar behave like the SM Higgs boson, etc. That such constraints are important 
can be seen from the lepton-specific 2HDM, where the requirement of large $\tan\beta$ forces large 
Yukawa couplings and destroys the stability of the potential at most above 1 TeV. However, it is too early
to say that such models are not compatible with both LHC data and solution of the naturalness problem; 
1 TeV is hardly a scale where one would talk about naturalness, and two-loop corrections might push the 
limit a bit upwards. 

We find solutions only for 
the large $\tan\beta$ case: about 31.5 for Type-II, 42.5 for flipped, and about 140 for lepton-specific 
models. In fact, the allowed values of $\tan\beta$ are severely restricted. The dependence of $\tan\beta$ on
$\lambda_1$ is almost linear for Type II and flipped 2HDM; 
for the lepton specific 2HDM the dependence is more complicated. Anyway, we believe that this 
 is an interesting result in the study of 2HDM. 

Not all the solutions for Type-II and flipped 2HDMs are valid till the Planck scale. 
In particular, whenever we 
start with a large quartic coupling, there is always a chance that the Landau pole will be hit before 
$M_{Pl}$. Fortunately, for most part of the parameter space, even if the blow-up occurs, the 
scale is rather large and beyond the reach of the LHC. Such a behaviour may be taken as indicative 
of some other new physics taking over at that scale. For lepton-specific models, the instability 
demands some new physics possibly within the LHC reach. 

Everything would have been perfect if the VCs were absolutely stable with the scale variation. 
Unfortunately, this is not so. But this may be due to the simplistic approach of keeping only the 
one-loop terms; at higher-orders, we might expect a scale independence.

 \section{Acknowledgements}

 The authors thank Debtosh Chowdhury for pointing out a mistake in the first version of the paper. 
 A.K.\ acknowledges DST, Government of India, and CSIR, Government of India, for research support. 
 I.C. acknowledges CSIR, Govt.\ of India, for a research fellowship.

\appendix
\numberwithin{equation}{section}

\section{RG equations}

Without the Yukawa couplings, the one-loop RGEs are 
 \bea
16\pi^2 \beta_{\lambda_1} &=& 6\lambda_1^2 + 2\lambda_3^2 + \lambda_4^2 + \lambda_5^2 + 
2\lambda_3\lambda_4  - \frac32\lambda_1 \left(g_1^2+3g_2^2\right) 
+ \frac{3}{8}(g_1^4 + 2g_1^2g_2^2 +3g_2^4)\,,\nonumber\\
16\pi^2\beta_{\lambda_2} &=& 6\lambda_2^2 + 2\lambda_3^2 + \lambda_4^2 + \lambda_5^2 + 
2\lambda_3\lambda_4  - \frac32\lambda_2 \left(g_1^2+3g_2^2\right)
 + \frac{3}{8}(g_1^4 + 2g_1^2g_2^2 +3g_2^4)\,,\nonumber\\
16\pi^2\beta_{\lambda_3} &=& (\lambda_1+\lambda_2)(3\lambda_3+\lambda_4) + 2 \lambda_3^2
+ \lambda_4^2 + \lambda_5^2 + \frac38 \left( 3g_2^4 + g_1^4 - 2 g_1^2 g_2^2 \right)- 
\frac32 \lambda_3\left(3g_2^2+g_1^2\right)\,,\nonumber\\ 
16\pi^2\beta_{\lambda_4} &=& \left(\lambda_1 + \lambda_2 + 4\lambda_3 + 2\lambda_4\right)\lambda_4 + 
4 \lambda_5^2 + \frac32g_1^2g_2^2 -\frac32\lambda_4\left(3g_2^2+ g_1^2\right)\,,\nonumber\\
16\pi^2\beta_{\lambda_5} &=& \left(\lambda_1 + \lambda_2 + 4 \lambda_3 + 6\lambda_4\right)\lambda_5 
- \frac32 \lambda_5\left(3g_2^2+g_1^2\right)\,.\nonumber\\
\label{rge1}
\eea

Once we include the Yukawa couplings, which are model-specific, the RGEs turn out to be as follows. 

\begin{itemize}
 \item Type II:
 \bea
 \left(16\pi^2 \beta_{\lambda_1}\right)_{total} &=& 16\pi^2 \beta_{\lambda_1} + 
 6\lambda_1\left(Y_1^b\right)^2 + 2\lambda_1\left(Y_1^{\tau}\right)^2
 - 6 \left(Y_1^b\right)^4 - 2 \left(Y_1^{\tau}\right)^4 \,,\nonumber\\
 \left(16\pi^2 \beta_{\lambda_2}\right)_{total} &=& 16\pi^2 \beta_{\lambda_2} + 
 6\lambda_2\left(Y_2^t\right)^2 
 - 6 \left(Y_2^t\right)^4 \,,\nonumber\\
 \left(16\pi^2 \beta_{\lambda_3}\right)_{total} &=& 16\pi^2 \beta_{\lambda_3} + 
 6\lambda_3\left(Y_1^b\right)^2 + 2\lambda_3\left(Y_1^{\tau}\right)^2
 +3\lambda_3\left(Y_2^t\right)^2 \,,\nonumber\\
 \left(16\pi^2 \beta_{\lambda_4}\right)_{total} &=& 16\pi^2 \beta_{\lambda_4} + 
 6\lambda_4\left(Y_1^b\right)^2 + 2\lambda_4\left(Y_1^{\tau}\right)^2
 +3\lambda_4\left(Y_2^t\right)^2 \,,\nonumber\\
 \left(16\pi^2 \beta_{\lambda_5}\right)_{total} &=& 16\pi^2 \beta_{\lambda_5} + 
 6\lambda_5\left(Y_1^b\right)^2 + 2\lambda_5\left(Y_1^{\tau}\right)^2
 +3\lambda_5\left(Y_2^t\right)^2 \,.\nonumber\\
 \label{rge2}
 \eea
 \item Lepton specific: 
 \bea
 \left(16\pi^2 \beta_{\lambda_1}\right)_{total} &=& 16\pi^2 \beta_{\lambda_1} + 
 2\lambda_1\left(Y_1^{\tau}\right)^2 - 2 \left(Y_1^{\tau}\right)^4 \,,\nonumber\\
 \left(16\pi^2 \beta_{\lambda_2}\right)_{total} &=& 16\pi^2 \beta_{\lambda_2} + 
 6\lambda_2\left(Y_2^t\right)^2 + 6\lambda_2\left(Y_2^b\right)^2
 - 6 \left(Y_2^t\right)^4 - 6 \left(Y_2^b\right)^4\,,\nonumber\\
 \left(16\pi^2 \beta_{\lambda_3}\right)_{total} &=& 16\pi^2 \beta_{\lambda_3} + 
 6\lambda_3\left(Y_2^b\right)^2 + 2\lambda_3\left(Y_1^{\tau}\right)^2
 +3\lambda_3\left(Y_2^t\right)^2 \,,\nonumber\\
 \left(16\pi^2 \beta_{\lambda_4}\right)_{total} &=& 16\pi^2 \beta_{\lambda_4} + 
 6\lambda_4\left(Y_2^b\right)^2 + 2\lambda_4\left(Y_1^{\tau}\right)^2
 +3\lambda_4\left(Y_2^t\right)^2 \,,\nonumber\\
 \left(16\pi^2 \beta_{\lambda_5}\right)_{total} &=& 16\pi^2 \beta_{\lambda_5} + 
 6\lambda_5\left(Y_2^b\right)^2 + 2\lambda_5\left(Y_1^{\tau}\right)^2
 +3\lambda_5\left(Y_2^t\right)^2 \,.\nonumber\\
\label{rge3}
 \eea
 \item Flipped: 
 \bea
 \left(16\pi^2 \beta_{\lambda_1}\right)_{total} &=& 16\pi^2 \beta_{\lambda_1} + 
 6\lambda_1\left(Y_1^b\right)^2 - 6 \left(Y_1^b\right)^4 \,,\nonumber\\
 \left(16\pi^2 \beta_{\lambda_2}\right)_{total} &=& 16\pi^2 \beta_{\lambda_2} + 
 6\lambda_2\left(Y_2^t\right)^2 + 2\lambda_2\left(Y_2^{\tau}\right)^2
 - 6 \left(Y_2^t\right)^4 - 2 \left(Y_2^{\tau}\right)^4\,,\nonumber\\
 \left(16\pi^2 \beta_{\lambda_3}\right)_{total} &=& 16\pi^2 \beta_{\lambda_3} + 
 3\lambda_3\left(Y_2^t\right)^2 + 2\lambda_3\left(Y_2^{\tau}\right)^2
 +6\lambda_3\left(Y_1^b\right)^2 \,,\nonumber\\
 \left(16\pi^2 \beta_{\lambda_4}\right)_{total} &=& 16\pi^2 \beta_{\lambda_4} + 
 3\lambda_4\left(Y_2^t\right)^2 + 2\lambda_4\left(Y_2^{\tau}\right)^2
 +6\lambda_4\left(Y_1^b\right)^2 \,,\nonumber\\
 \left(16\pi^2 \beta_{\lambda_5}\right)_{total} &=& 16\pi^2 \beta_{\lambda_5} + 
 3\lambda_5\left(Y_2^t\right)^2 + 2\lambda_5\left(Y_2^{\tau}\right)^2
 +6\lambda_5\left(Y_1^b\right)^2 \,.\nonumber\\
 \label{rge4}
 \eea
\end{itemize}
Note that we have not talked about the RGEs of Type-I 2HDM, the reasons are given in the 
main body of the paper. 

The Yukawa RGEs are:
\begin{itemize}

\item Type-II:
\bea
 16\pi^2 \beta_{Y_1^b} &=& \frac12 Y_1^{b}\left[-8g_s^2 - \frac94 g_2^2 - \frac{5}{12} g_1^2+
 \frac92\left( Y_1^{b}\right)^2 + \left( Y_1^{\tau}\right)^2 + \frac12\left( Y_2^{t}\right)^2\right]\,\nonumber\\
 16\pi^2 \beta_{Y_2^t} &=& \frac12 Y_2^{t}\left[-8g_s^2 - \frac94 g_2^2 - \frac{17}{12} g_1^2+
 \frac92\left( Y_2^{t}\right)^2 + \frac12\left( Y_1^{b}\right)^2\right]\,\nonumber\\
 16\pi^2 \beta_{Y_1^\tau} &=& \frac12 Y_1^{\tau}\left[-\frac94 g_2^2 - \frac{15}{4} g_1^2+
 3\left( Y_1^{b}\right)^2 + \frac52\left( Y_1^{\tau}\right)^2\right]\,\nonumber\\
 \eea
 
\item Lepton specific: 
 \bea
 16\pi^2 \beta_{Y_2^b} &=& \frac12 Y_2^{b}\left[-8g_s^2 - \frac94 g_2^2 - \frac{5}{12} g_1^2+
 \frac92\left( Y_2^{b}\right)^2 + \frac32\left( Y_2^{t}\right)^2\right]\,\nonumber\\
 16\pi^2 \beta_{Y_2^t} &=& \frac12 Y_2^{t}\left[-8g_s^2 - \frac94 g_2^2 - \frac{17}{12} g_1^2+
 \frac92\left( Y_2^{t}\right)^2 + \frac32\left( Y_2^{b}\right)^2\right]\,\nonumber\\
 16\pi^2 \beta_{Y_1^\tau} &=& \frac12 Y_1^\tau\left[-\frac94 g_2^2 - \frac{15}{4} g_1^2+
 \frac52\left( Y_1^\tau\right)^2\right]\,\nonumber\\
 \eea
 
\item Flipped:
\bea
 16\pi^2 \beta_{Y_1^b} &=& \frac12 Y_1^b\left[-8g_s^2 - \frac94 g_2^2 - \frac{5}{12} g_1^2+
 \frac92\left( Y_1^b\right)^2 + \frac12\left( Y_2^{t}\right)^2\right]\,\nonumber\\
 16\pi^2 \beta_{Y_2^t} &=& \frac12 Y_2^{t}\left[-8g_s^2 - \frac94 g_2^2 - \frac{17}{12} g_1^2+
 \frac92\left( Y_2^t\right)^2 + \frac12\left( Y_1^{b}\right)^2+ \left( Y_2^{\tau}\right)^2\right]\,\nonumber\\
 16\pi^2 \beta_{Y_2^\tau} &=& \frac12 Y_2^{\tau}\left[-\frac94 g_2^2 - \frac{15}{4} g_1^2+
 \frac52\left( Y_2^\tau\right)^2 + 3 \left( Y_2^{t}\right)^2\right]\,\nonumber\\
 \eea
\end{itemize} 

 Here $g_s$, $g_1$ and $g_2$ are gauge coupling constants of ${SU(3)}_c$ , $U(1)_Y$ and $SU(2)_L$ respectively,
$\beta_h\equiv dh/dt$, and $t \equiv \ln(q^2/\mu^2)$.  Note that our definition of $t$ differs 
by a factor of 2 from that used by some authors.

A more compact form of the one-loop RGEs for the Yukawa couplings, valid for all 
2HDMS, is \cite{Branco:PhysRept}
\bea
16\pi^2 \beta_{Y_j^b} &=& \frac12 a_b Y_j^b + \frac12 \sum_{k=1}^{n_H} T_{jk} Y_k^b \nonumber\\
&&  +\frac12 \sum_{k=1}^{n_H} \left(-2 Y_k^t {Y_j^t}^{\dag}Y_k^b+\frac12 Y_k^t{Y_k^t}^{\dag}Y_j^b + Y_j^b {Y_k^b}^{\dag}Y_k^b + \frac12 Y_k^b{Y_k^b}^{\dag}Y_j^b\right)\,,
\eea
\bea
16\pi^2 \beta_{Y_j^t} &=& \frac12 a_t Y_j^t + \frac12 \sum_{k=1}^{n_H}T_{jk}^* Y_k^t \nonumber\\
&&  +\frac12 \sum_{k=1}^{n_H} \left(-2 Y_k^b {Y_j^b}^{\dag}Y_k^t+\frac12 Y_k^b{Y_k^b}^{\dag}Y_j^t + Y_j^t {Y_k^t}^{\dag}Y_k^t + \frac12 Y_k^t{Y_k^t}^{\dag}Y_j^t\right)\,,
\eea
\be
16\pi^2 \beta_{Y_j^\tau} = \frac12 a_{\tau} Y_j^{\tau} + \frac12 \sum_{k=1}^{n_H} T_{jk} Y_k^{\tau} 
+\frac12 \sum_{k=1}^{n_H} \left(\frac12 Y_k^{\tau}{Y_k^{\tau}}^{\dag}Y_j^{\tau} + Y_j^{\tau} {Y_k^{\tau}}^{\dag}Y_k^{\tau}\right)\,,
\ee
where
\bea
a_b &=& -8g_s^2 - \frac 94 g_2^2 - \frac {5}{12} g_1^2\,,\nonumber\\
a_t &=& -8g_s^2 - \frac 94 g_2^2 - \frac {17}{12} g_1^2\,,\nonumber\\
a_{\tau} &=& - \frac 94 g_2^2 - \frac {15}{4} g_1^2\,,\nonumber\\
T_{jk} &=& 3 tr\left(Y_j^b {Y_k^b}^{\dag} + {Y_j^t}^{\dag} Y_k^t\right) + tr\left(Y_j^{\tau} {Y_k^{\tau}}^{\dag}\right)\,.
\eea

\end{document}